\documentclass[submission,copyright,creativecommons]{eptcs}
\providecommand{\event}{DCM 2014} % Name of the event you are submitting to
\usepackage{etex}
\usepackage{graphicx}
\usepackage[disable,colorinlistoftodos]{todonotes}
\usepackage{tabularx}
\usepackage{bussproofs}
\usepackage{proof}
\usepackage{xypic}
\usepackage{fancyhdr}
\usepackage{pdflscape}

\usepackage[english]{babel}
\usepackage{url}
\usepackage{algorithm}
\usepackage{algorithmic}
\usepackage{amsmath} 
\usepackage{amssymb}
\usepackage{color}

\newtheorem{definition}{Definition}
\newtheorem{theorem}{Theorem}
\newtheorem{corollary}{Corollary}

\newtheorem{lemma}{Lemma}
\newtheorem{proposition}{Proposition}

\newcommand{\Log}[1]{\ensuremath{\mathcal{#1}}}
\newcommand{\NatDed}[1]{\ensuremath{\mathcal{N}_{\mathcal{#1}}}}

\newcommand{\limply}{\supset}
\newcommand{\bm}[1]{\ensuremath{\mathbf{#1}}}

\newcommand{\imply}{\ensuremath{\rightarrow}}

\newcommand{\mil}{\ensuremath{\mathbf{M}_{\rightarrow}~}}

\newcommand{\SEQ}{\ensuremath{\Rightarrow}}

\def\fCenter{ \SEQ\ }

\title{Propositional logics complexity and the sub-formula property}
\author{Edward Hermann  Haeusler \institute{PUC-Rio \\ Rio de Janeiro, Brasil}
\email{hermann@inf.puc-rio.br}}
\def\titlerunning{Propositional logics complexity}
\def\authorrunning{E.H. Haeusler}

\begin{document}

\maketitle

 \begin{abstract}
 In 1979 Richard Statman proved, using proof-theory, that the purely implicational fragment of Intuitionistic Logic (\mil) is PSPACE-complete. He showed a polynomially bounded translation from full Intuitionistic Propositional Logic into its implicational fragment. By the PSPACE-completeness of S4, proved by Ladner, and the G\oe del translation from S4 into Intuitionistic Logic, the PSPACE-completeness of \mil~ is drawn. 
The sub-formula principle for a deductive system for a logic \Log{L} states that whenever $\{\gamma_1,\ldots,\gamma_k\}\vdash_{\Log{L}}\alpha$ there is a proof in which each formula occurrence is either a sub-formula of $\alpha$ or of some of $\gamma_i$.  In this work we extend Statman's result and show that any propositional (possibly modal) structural logic satisfying a particular formulation of the sub-formula principle is in PSPACE. If the logic includes the minimal purely implicational logic then it is PSPACE-complete. 
 As a consequence, EXPTIME-complete propositional logics, such as PDL and the common-knowledge epistemic logic with at least 2 agents satisfy this particular sub-formula principle, if and only if, PSPACE=EXPTIME. We also show how our technique can be used to prove that any finitely many-valued logic has the set of its tautologies in $PSPACE$.
 \end{abstract}

\section{Introduction}\label{sec:ialc}

In \cite{Statman}, R. Statman showed a polynomial-time reduction from Intuitionistic Propositional Logic into its implicational fragment. This reduction proves that Purely Implicational Minimal Logic is PSPACE-complete. The methods that Statman uses in \cite{Statman} are based on proof-theory and Natural Deduction in Prawitz Style. The sub-formula principle for a Natural Deduction system \NatDed{L} for a logic \Log{L} states that whenever $\alpha$ is provable from $\Gamma$, in \Log{L}, there is a  derivation of $\alpha$ from a set of assumptions $\{\delta_1,\ldots,\delta_k\}\subseteq\Gamma$ built up only with sub-formulas of $\alpha$ and/or $\{\delta_1,\ldots,\delta_k\}$. In this article we show that the Validity problem for any propositional logic \Log{L}, with a Natural Deduction system that satisfies the sub-formula principle, is in PSPACE. Besides that, if \Log{L} includes the usual $\imply$-rules then  PSPACE-complete.  

In \cite{Pra:promcl}, we can found a general approach to Natural Deduction that allows the definition of general introduction and elimination rules in a way that, any intuitionistic logical constant (sometimes called operator or connective) that is expressible in terms of these general rules is also expressible by means of the intuitionistic logical constants ($\bot, \imply, \land, \neg$ and $\lor$). Precisely, in \cite{Pra:promcl} it is shown that any constant \bm{c} determined by a set of introduction and elimination rules, in Natural Deduction style, is such that, there is a formula $F(A_1,\ldots,A_k)$ built up with constants from $\{\bot, \imply, \land, \neg,\lor\}$, and, $\bm{c}(A_1,\ldots,A_k)$ is provable, iff, $F(A_1,\ldots,A_k)$ is provable in intuitionistic propositional logic. 
In order to prove this statement, namely, the functional completeness of intuitionistic logical constants, some proof-theoretical assumptions on the relationship between elimination and introduction rules are considered. Precisely, these assumptions have to do with the inversion principle that roughly states that an elimination rule for a logical constant \bm{c} has to be a function of respective introduction rules of this \bm{c}. This means that the elimination rule is only determined by the introduction rules and this inversion principle\footnote{A good discussion on the form of the inversion principle can be found in \cite{StephenRead}}. 

Although strongly based on \cite{Pra:promcl}, the  results shown here,  do not explicitly need inversion principle. We do use a general form of introduction and elimination rules proposed in \cite{Pra:promcl} and after extended by Roy Dickhoff and Nissim Francez to the term {\em general-elimination harmony}, such that, there is only one \bm{c}-elimination rule associated to the many \bm{c}-introduction rules. As we see in Section~\ref{translation}, this {\em general-elimination harmony} assumption is not needed, indeed. In \cite{StephenRead} there is a very good discussion on the Higher-Level rules, proposed in \cite{Schroeder-Heister} as an alternative and extension of Prawitz \cite{Pra:promcl}, and the {\em general-elimination harmony} as well. To sum it up , our technique relies in the sub-formula property only, as it is stated in the Definition~\ref{SFP} in the following section. We have to mention that \cite{Arnon} also provides an approach that does not require a full harmony between introduction and elimination rules, but \cite{Arnon} does not have the same goal as ours, in providing a computational complexity analysis based on proof-theoretical arguments as ours.    

%% Section~\ref{translation} shows our main result and in section~\ref{applications} we discuss some consequences and applications of our main result. At the conclusion we discuss further possible investigations. 

\section{Translating a propositional logic $\mathcal{L}$ into \mil}\label{translation}

Consider a propositional logic $\mathcal{L}$ with Natural Deduction system having introduction and elimination rules according to the following general schema.  A  $\mathbf{c}$-introduction rule, as shown in Figure~\ref{general-intro-rules}, derives $\mathbf{c}(\beta_1,\ldots,\beta_n,\phi^{1}_1,\ldots,\phi^{1}_{j_1},\ldots,\phi^{n}_1,\ldots,\phi^{n}_{j_1},\gamma_{1},\ldots,\gamma_{m})$, discharging occurrences of the bracket formulas $[\phi^{i}_1],\ldots,[\phi^{i}_{j_i}]$ in the respective derivation of $\beta_i$.  For reasons of readability, we  sometimes write $\bm{c}(\beta_i,\phi^{j},\gamma_{k})$
as a shortened (abbreviation)  of $\mathbf{c}(\beta_1,\ldots,\beta_n,\phi^{1}_1,\ldots,\phi^{1}_{j_1},\ldots,\phi^{n}_1,\ldots,\phi^{n}_{j_1},\gamma_{1},\ldots,\gamma_{m})$\footnote{Consider usual $\lor$-intro rules. Both have $\beta_1\lor\beta_2$ as conclusion, one has $\beta_1$ as premise and the other has $\beta_2$ as the only premise. According to our schema, $\beta_2$ plays the role of $\gamma_1$ in the first rule, while $\beta_1$ plays this role in the second $\lor$-intro rule}. An elimination rule, corresponding to the same operator $\mathbf{c}$ is shown in Figure~\ref{general-elim-rules} below. It is important to say that the discharging discipline is liberal, that is, when applying a rule, it is possible to discharge multiple assumptions including vacuous discharges. In this way, {\em this general schema cannot be used with sub-structural logics like Relevance and Linear Logics.} It is also important to note that each $\phi^{i}_{j}$ determine the type of the formula that can be discharged, not an instance of formula. For example, when representing the $\imply$-introduction rule, there is only one type $\phi^{1}_{1}$and  $c(\beta_1,\phi^{1}_{1})$ is  $\phi^{1}_{1}\imply\beta_1$. The $\imply$-intro rule is shown in Figure~\ref{introimply} below as an instance of the general schema.

\begin{figure}[h]
\begin{prooftree}
\AxiomC{$[\phi^{1}_1],\ldots,[\phi^{1}_{j_1}]$}
\noLine
\UnaryInfC{$\mid$}
\noLine
\UnaryInfC{$\beta_1$}
\AxiomC{$\ldots$}
\AxiomC{$[\phi^{n}_1],\ldots,[\phi^{n}_{j_n}]$}
\noLine
\UnaryInfC{$\mid$}
\noLine
\UnaryInfC{$\beta_n$}
\TrinaryInfC{$\mathbf{c}(\beta_1,\ldots,\beta_n,\phi^{1}_1,\ldots,\phi^{1}_{j_1},\ldots,\phi^{n}_1,\ldots,\phi^{n}_{j_1},\gamma_{1},\ldots,\gamma_{m})$}
\end{prooftree}
\caption{$\bm{c}$-introduction rule schema}\label{general-intro-rules}
\end{figure}

\begin{figure}[h]
\begin{prooftree}
\AxiomC{$\bm{c}(\beta_i,\phi^{j} )\;\;\;(\phi^{j}_i)_{i,j}$}
\AxiomC{$[\beta_1]$}
\noLine
\UnaryInfC{$\;\;\;\mid\ldots$}
\noLine
\UnaryInfC{$\chi$}
%\AxiomC{$\ldots$}
\AxiomC{$[\beta_n]$}
\noLine
\UnaryInfC{$\ldots\mid\;\;\;$}
\noLine
\UnaryInfC{$\chi$}
\TrinaryInfC{$\chi$}
\end{prooftree}
\caption{$\bm{c}$-elimination rules schemata}\label{general-elim-rules}
\end{figure}

\begin{figure}[h]
\begin{prooftree}
\AxiomC{$[\phi^{1}_1]$}
\noLine
\UnaryInfC{$\mid$}
\noLine
\UnaryInfC{$\beta_1$}
\UnaryInfC{$\phi^{1}_{1}\imply\beta_1$}
\end{prooftree}
\caption{$\imply$-introduction as an instance of the general schema}\label{introimply}
\end{figure}

In elimination rules (see Figure~\ref{general-elim-rules}), $\phi^{j}_i$, $i=1,j_i$, are called simple-minor premise, for each $j$. The premises $\chi$ are called discharging minor premises (d-minor premise). The premise $\bm{c}(\beta_i,\phi^{j} )$ is the major premise of the elimination rule. 
There may be more than one introduction rule, but at most one elimination rule. This elimination rule has one discharging minor premise for each introduction rule. As a matter of illustration we show below how the usual $\imply$-elim rule is seen  as instance of a general elimination schema. In order to see that, remember that the usual (minimal) implication has only one introduction rule that can be easily seen as an instance of an introduction schema. The elimination schema consider this instance of introduction schema, resulting in the shown rule. The usual form of the $\imply$-elim is obtained by proving $B$ instead of discharging it. For a more detailed explanation on these general rules, we recommend \cite{Pra:promcl}. 

%\begin{figure}[h]
\def\proofSkipAmount{\vskip .7cm}
\begin{prooftree}
\AxiomC{$A\imply B$}
\AxiomC{$A$}
\AxiomC{$[B]$}
\noLine
\UnaryInfC{$\mid$}
\noLine
\UnaryInfC{$C$}
\TrinaryInfC{$C$}
%\LabelRight{$\imply$-Elim}
\end{prooftree}
%\caption{$\imply$-elim rule from general elimination schema}\label{imply-rules}
%\end{figure}

Besides intuitionistic logic, some modal classical logics have Natural Deduction systems that conforms with our schema. For example, the Natural Deduction system presented in \cite{Pra:ND}, page 75,  for the logics S4 and S5 can be viewed as instances of our schemata, considering additional provisos for ensuring the soundness of the respective system. The rule for the $\diamond$-introduction has no proviso and is as follows.
\def\proofSkipAmount{\vskip .7cm}
\begin{prooftree}
\AxiomC{$A$}
\UnaryInfC{$\diamond A$}
\end{prooftree} 
The $\diamond$-elimination rule has a proviso, stated below, and it is as follows:
\def\proofSkipAmount{\vskip .7cm}
\begin{prooftree}
\AxiomC{$\diamond A$}
\AxiomC{$A$}
\noLine
\UnaryInfC{$\mid$}
\noLine
\UnaryInfC{$B$}
\BinaryInfC{$B$}
\end{prooftree}
where: (1) For S4, in the proof of the minor premise $B$, $B$ only depends on formulas $\Box\alpha$ or $\neg\diamond\alpha$, but the formula occurrences $A$ discharged by the rule  (2) For S5, in the proof of the minor premise $B$, $B$ only depends on formulas $\Box\alpha$, $\diamond\alpha$,$\neg\diamond\alpha$, $\neg\Box\alpha$, but the formula occurrences $A$ discharged by the rule. 

In abstract proof-theory, we would require that the principle of inversion holds concerning the elimination and introduction rules. This usually is among the requirements to have a normalization theorem holding for a Natural Deduction system. However, for our analysis, nothing is required but the sub-formula principle, defined in Definition~\ref{SFP}.

A logical constant $\bm{c}$, also called {\em operator},  can be a propositional connective or a modality. A Natural Deduction system for a logic \Log{L} is a set of introduction and/or elimination rules for the operators of \Log{L}. Given a Natural Deduction system \NatDed{L} for a logic \Log, the usual notions of derivation and proof of a formula from a set of assumptions are considered in this article.  It is worth observing the difference between formula and formula occurrence.

\begin{definition}[Linked formulas and sequences]\label{branch}
Let \NatDed{L} be a Natural Deduction for \Log{L}. Let $\Pi$ be a derivation of a formula $\alpha$ from a set of assumptions $\Gamma$ in \NatDed{L}.
Let $\beta_1$ and $\beta_2$ be two formula occurrences in $\Pi$. We say that $\beta_1$ is linked to $\beta_2$, if and only if, there is a sequence $\delta_0,\ldots,\delta_k$ of formula occurrences in $\Pi$, such that:
%\begin{itemize}
(1) $\delta_0$ is $\beta_1$ and $\delta_k$ is $\beta_2$, and, for each $i=0,\ldots,k-1$;
(2) $\delta_i$ is major premise of an application of a $\bm{c}$-elim  rule, in $\Pi$,  and $\delta_{i+1}$ is a formula occurrence discharged by this application of elimination rule, or;
(3) $\delta_i$ is a discharging minor premise of an application of a  $\bm{c}$-elim rule, in $\Pi$,  and $\delta_{i+1}$ is its conclusion, or;
(4) $\delta_i$ is a premise of an application of a  $\bm{c}$-intro rule, in $\Pi$,  and $\delta_{i+1}$ is its conclusion.
%\end{itemize}
\end{definition}

We call $\delta_0,\ldots,\delta_k$ a linking-sequence in $\Pi$. 
Observing the definition above, we note that a formula occurrence that is either a simple-minor premise of an application of a $\bm{c}$-elim rule in $\Pi$ or the conclusion of $\Pi$ cannot be linked to any other formula occurrence in $\Pi$ any more. Thus, the longest possible linking-sequences in any derivation $\Pi$ have $\delta_k$ as a simple-minor premise or the conclusion of $\Pi$. Our definition of linking-sequence is similar, and generalizes, the definition of branches as stated in \cite{Pra:ND}. 

\begin{definition}
Let $\Pi$ be a derivation in \NatDed{L}. Consider a linking-sequence $\delta_0,\ldots,\delta_k$ in $\Pi$. We say that $\delta_0,\ldots,\delta_k$ is a sub-formula linking-sequence, if and only if, for every $i=0,\ldots,k$, either $\delta_i$ is sub-formula of $\delta_{i+1}$, or, $\delta_{i+1}$ is sub-formula of $\delta_{i}$.
\end{definition}

\begin{definition}[Derivations satisfying the Sub-Formula Property]
Let $\mathcal{L}$ be a logic and $\mathcal{N}_{\mathcal{L}}$ a Natural Deduction system for $\mathcal{L}$. Let $\Pi$ be a derivation of $\alpha$ from $\{\delta_1,\ldots,\delta_k\}$ in  $\mathcal{N}_{\mathcal{L}}$. We say that $\Pi$ satisfy the sub-formula principle, if and only if, every linking-sequence in $\Pi$ is a sub-formula linking-sequence and every formula occurring in this linking-sequence is either a  sub-formula of $\alpha$ or sub-formula of some of the formulas in $\Gamma$. 
\end{definition}

\begin{definition}[Systems satisfying the Sub-Formula Property]\label{SFP}
We say that a system \NatDed{L} satisfy the sub-formula property, if and only if, for every derivation $\Pi$ of $\alpha$ from $\Gamma$ in it, there is a derivation $\Pi^{\prime}$  of $\alpha$ from $\Gamma^{\prime}\subseteq\Gamma$ satisfying the sub-formula property. 
\end{definition}

When a logic \Log{L} has a system \NatDed{L} satisfying the sub-formula property, we say that the logic \Log{L} itself satisfies the sub-formula property. Thus, it is important to emphasize that to satisfy the sub-formula property means that the logic has a Natural Deduction system according the general rules shown here. This also means that sub-structural logics do not satisfy our notion of sub-formula property. 

\vspace{1cm}

\begin{proposition}\label{sub-formula}
Let $\Pi$ be a derivation satisfying the sub-formula principle. If $\Pi^{\prime}$ is a sub-derivation of $\Pi$ then $\Pi^{\prime}$ satisfies the sub-formula principle too. 
\end{proposition}

{\em Proof of proposition}. We observe that a sub-sequence of a linking-sequence that is a sub-formula linking-sequence is a sub-formula linking-sequence too. Since every linking-sequence of $\Pi$ is a sub-formula linking sequence,  then every linking sequence of $\Pi^{\prime}$ is a linking sequence too. 
\begin{center}
Q.E.D.
\end{center}

In what follows we define a translation $\star$ from $\mathcal{L}$ into \mil, such that, $\alpha$ is provable in $\mathcal{L}$, if and only if,  $\alpha^{\star}$ is provable in \mil. In order to define $\star$, we need auxiliary functions $\mathcal{M}$ and $\mathcal{A}$, defined in the sequel. In the following definition, $p_{\omega}$ is a notation used to uniquely identify the symbol $p$ indexed by the string (word) $\omega$. It this way we ensure that $p$ is the unique propositional symbol that is named by means of the word $\omega$, such that,  $p_{\omega_1}$ and $p_{\omega_2}$ are equal, if and only if, $\omega_1$ and $\omega_2$ are the same word (string). We remember that we call operator any propositional connective or modality. We consider that the logic $\mathcal{L}$ has a finite set $\{\bm{c_1},\ldots,\bm{c_k}\}$ of operators. For reasons of readability, we do not consider the minimal implication $\imply$ in this set, even in the case that the logic has it as one of its propositional connectives. As we will see, the implication is the only logical constant (operator) that is not translated. So, in what follows we consider only logics including the purely minimal implicational logic.  

The propositional logics considered in this article have in their Natural Deduction systems the usual $\imply$-intro and $\imply$-elim rules.

The following definitions provide us axioms schemata  concerning each $\bm{c}$-introduction and/or $\bm{c}$-elimination rule.

\begin{definition}[$\iota$-axiom]
Consider an introduction rule $r$  for an operator $\bm{c}$ as shown in Figure~\ref{general-intro-rules}. The $\iota$-axiom concerning this rule schema $r$ instantiated to formulas $\beta_i$ and $\phi^{i}_{1},\ldots,\phi^{i}_{j_i}$, denoted by $\iota(r,\beta_i,\phi^{j},\gamma_{k})$, is the following implicational formula:
\[
(\phi^{1}_1\imply(\ldots\imply (\phi^{1}_{j_1}\imply\beta_1)))\imply\ldots(\phi^{n}_1\imply(\ldots\imply (\phi^{n}_{j_n}\imply\beta_n)))\imply p_{\bm{c}(\beta_i,\phi^{j},\gamma_k)})
\]
\end{definition}  

\begin{definition}[$\epsilon$-axiom]
Consider a formula $\chi$ of a logic $\mathcal{L}$ and an elimination rule $r$ for $\bm{c}$, as shown right in Figure~\ref{general-elim-rules} instantiated to $\phi^{j}$, $\beta_i$ and $\chi$. The $\epsilon$-axiom concerning this rule schema and $\chi$, denoted by $\epsilon(r,\chi,\beta_i,\phi^{j})$, is the following implicational formula:
\[
\phi^{1}_{1}\imply(\ldots(\phi^{n}_{j_n}\imply(..(\beta_1\imply\chi)\imply\ldots(\beta_n\imply\chi)\imply(p_{\bm{c}(\beta_i,\phi^{j})}\imply\chi))))
\]
\end{definition} 

For example, considering the usual $\lor$-introduction ($\lor-intro_1$ and $\lor-intro_2$) and $\lor$-elimination ($\lor-elim$) natural deduction schemata, we have that: 
(1) $\iota(\lor-intro_1,\beta_1,\beta_2)$ is $\beta_1\imply p_{\lor(\beta_1,\beta_2)}$; (2) $\iota(\lor-intro_2,\beta_1,\beta_2)$ is $\beta_2\imply p_{\lor(\beta_1,\beta_2)}$, and; (3) $\epsilon(\lor-elim,\beta_1,\beta_2,\chi)$ is 
$(\beta_1\imply \chi)\imply((\beta_2\imply\chi)\imply (p_{\lor(\beta_1,\beta_2)}\imply\chi))$.  

\begin{definition}[Atomizing Operators]
The mapping $\mathcal{M}$ from the language of $\mathcal{L}$ into the one of \mil is defined inductively,  as follows: 
%\begin{description}
{\bf Atoms} $\mathcal{M}(p) = p$, if $p$ is a propositional letter;
{\bf Implication} $\mathcal{M}(\alpha_1\imply\alpha_2) = \mathcal{M}(\alpha_1)\imply\mathcal{M}(\alpha_2)$;
{\bf Operators} $\mathcal{M}(\bm{c_m}(\beta_i,\phi^{j},\gamma_k )) = p_{\bm{c_m}(\beta_i,\phi^{j},\gamma_k )}$, if $\bm{c_m}$ is an operator of $\mathcal{L}$.
%\end{description}
\end{definition}

The second clause in the definition above is only used when the language of $\mathcal{L}$ includes the minimal implication $\imply$. Otherwise it is not used and the translation is also well-defined. In the following we define an auxiliary function that for each formula $\alpha$ yields the set of implicational formulas that express the ``deductive meaning''\footnote{This informal expression is made precise in the statement of Theorem~\ref{principal}} of the elimination and introduction rules for each operator in $\mathcal{L}$.

\vspace{1cm}

\begin{definition}[Axiomatizing Operators]
Given a formula $\alpha$ in $\mathcal{L}$, 
the mapping $\mathcal{A}^{\alpha}$ from the language of $\mathcal{L}$ into (finite) sets of formulas in the language of \mil is defined inductively,  as follows.
\begin{description}
\item[$\bullet$ Atoms] $\mathcal{A}^{\alpha}(p) = \emptyset$;
\item[$\bullet$ Implication] $\mathcal{A}^{\alpha}(\alpha_1\imply\alpha_2)=\mathcal{A}^{\alpha}(\alpha_1)\cup\mathcal{A}^{\alpha}(\alpha_2)$;
\item[$\bullet$ Operators] $\mathcal{A}^{\alpha}(\bm{c_m}(\beta_{i},\phi^{j},\gamma_k))=$
\[
\begin{array}{c}
\{\iota(r,\mathcal{M}(\beta_{i}),\mathcal{M}(\phi^{j}),\mathcal{M}(\gamma_k))/\mbox{$r$ is a $\bm{c_m}$-intro rule}\} 
\\ \cup \\ 
\{\epsilon(r,\mathcal{M}(\chi),\mathcal{M}(\beta_{i}),\mathcal{M}(\phi^{j}))/\mbox{$r$ is a $\bm{c_m}$-elim rule and $\chi\in sub(\alpha)$}\}
\end{array}
\]
\end{description}
\end{definition}
\vspace{1cm}

\begin{lemma}\label{lemma1}
Let $\mathcal{L}$ be a logic having a Natural Deduction system satisfying sub-formula property. Let $\Pi$ be a proof of $\alpha$ from $\Gamma$ in $\mathcal{L}$. There is a derivation $\Pi^{\prime}$ of $\mathcal{M}(\alpha)$ from $\Gamma^{\prime}\subseteq \mathcal{M}(\Gamma)\cup\bigcup_{\gamma\in\Gamma}\mathcal{A}^{\gamma}(\Gamma,\alpha)\cup\mathcal{A}^{\alpha}(\Gamma,\alpha)$ in \mil. We use the notation $\mathcal{A}^{\alpha}(\Gamma)$ to denote $\bigcup_{\gamma\in\Gamma}\mathcal{A}^{\alpha}(\gamma)$ and ``$\Gamma,\alpha$''  to denote $\Gamma\cup\{\alpha\}$. 
\end{lemma}

{\em Proof of Lemma.} Since $\mathcal{L}$ satisfies the sub-formula principle, we can consider that $\Pi$ is a derivation satisfying the sub-formula principle. The proof proceeds by 
induction on the size of this derivation.  The basis is the derivation of $\alpha$ from $\alpha$ itself, and hence, $\Pi^{\prime}$ is $\mathcal{M}(\alpha)$ only. The inductive step is according the last rule applied in the derivation satisfying the sub-formula principle. 
There are only two (general) cases:
\begin{description}
\item[$\bullet$ Last rule is a $\bm{c_m}$-intro rule.] Then $\Pi$ is as following:
\def\proofSkipAmount{\vskip .8cm}
\begin{prooftree}
\AxiomC{$[\phi^{1}_1],\ldots,[\phi^{1}_{j_1}]$}
\noLine
\UnaryInfC{$\Pi^{\star}_1$}
\noLine
\UnaryInfC{$\beta_1$}
\AxiomC{$\ldots$}
\AxiomC{$[\phi^{n}_1],\ldots,[\phi^{n}_{j_n}]$}
\noLine
\UnaryInfC{$\Pi^{\star}_n$}
\noLine
\UnaryInfC{$\beta_n$}
\TrinaryInfC{$\mathbf{c}(\beta_1,\ldots,\beta_n,\phi^{1}_1,\ldots,\phi^{1}_{j_1},\ldots,\phi^{n}_1,\ldots,\phi^{n}_{j_1},\gamma_{1},\ldots,\gamma_{m})$}
\end{prooftree}
Where the last rule is $r$, a $\bm{c}$-introduction rule. Consider $\iota(r,\beta_i,\phi^{j},\gamma_k)$ the implicational formula related to $\bm{c}$.  
By Proposition~\ref{sub-formula} $\Pi^{\star}_{i}$ satisfies the sub-formula principle,  $i=1,n$. By inductive hypothesis, for each $i=1,n$, there is a derivation $\Pi^{\prime}_{i}$ satisfying the statement of the lemma.
Each $\Pi^{\prime}_{i}$ is of the following form.
\def\proofSkipAmount{\vskip .8cm}
\begin{prooftree}
\AxiomC{$[\mathcal{M}(\phi^{i}_1)],\ldots,[\mathcal{M}(\phi^{i}_{j_i})]$}
\noLine
\UnaryInfC{$\Pi^{\prime}_i$}
\noLine
\UnaryInfC{$\mathcal{M}(\beta_i)$}\LeftLabel{$\imply$-i}
\end{prooftree}
Thus, for each $i$ we can build, from $\Pi^{\prime}_i$, a derivation as the following, which we will denote by $\Sigma_i$. 
\def\proofSkipAmount{\vskip .8cm}
\begin{prooftree}
\AxiomC{$[\mathcal{M}(\phi^{i}_1)],\ldots,[\mathcal{M}(\phi^{i}_{j_i})]$}
\noLine
\UnaryInfC{$\Pi^{\prime}_i$}
\noLine
\UnaryInfC{$\mathcal{M}(\beta_i)$}\LeftLabel{$\imply$-i}
%\doubleLine
\UnaryInfC{$\mathcal{M}(\phi^{i}_{j_i})\imply\mathcal{M}(\beta_i)$}
\UnaryInfC{$\mathcal{M}(\phi^{i}_{j_i-1})\imply(\mathcal{M}(\phi^{i}_{j_i})\imply\mathcal{M}(\beta_i))$}
\noLine
\UnaryInfC{$\vdots$}
\noLine
\UnaryInfC{$(\mathcal{M}(\phi^{i}_1)\imply(\ldots\imply(\mathcal{M}(\phi^{i}_{j_i})\imply\mathcal{M}(\beta_i))..))$}
\end{prooftree}
Finally, the derivation below is a derivation in \mil of the form stated by the lemma, that is, of $\mathcal{M}(\mathbf{c}(\beta_1,\ldots,\beta_n,\phi^{1}_1,\ldots,\phi^{1}_{j_1},\ldots,\phi^{n}_1,\ldots,\phi^{n}_{j_1},\gamma_{1},\ldots,\gamma_{m}))$ = $p_{\bm{c}(\beta_i,\phi^{j},\gamma_k)}$. For reasons of space we abbreviate $(\mathcal{M}(\phi^{i}_1)\imply(\ldots\imply(\mathcal{M}(\phi^{i}_{j_i})\imply\mathcal{M}(\beta_i))..))$ by $B_i$.  
\def\proofSkipAmount{\vskip 1cm}
\begin{prooftree}
\AxiomC{$[\mathcal{M}(\phi^{n}_1)],\ldots,[\mathcal{M}(\phi^{n}_{j_n})]$}
\noLine
\UnaryInfC{$\Sigma_n$}
\noLine
\UnaryInfC{$B_n$}%\LeftLabel{$\imply$-i}

\AxiomC{$[\mathcal{M}(\phi^{2}_1)],\ldots,[\mathcal{M}(\phi^{2}_{j_2})]$}
\noLine
\UnaryInfC{$\Sigma_{2}$}
\noLine
\UnaryInfC{$B_2$}%\LeftLabel{$\imply$-i}
\AxiomC{$[\mathcal{M}(\phi^{1}_1)],\ldots,[\mathcal{M}(\phi^{1}_{j_1})]$}
\noLine
\UnaryInfC{$\Sigma_1$}
\noLine
\UnaryInfC{$B_1$}%\LeftLabel{$\imply$-i}

\AxiomC{$\iota(r,\mathcal{M}(\beta_i),\mathcal{M}(\phi^{j}),\mathcal{M}(\gamma_k))$}

\BinaryInfC{$B_2\imply(B_3\imply\ldots(B_n\imply p_{\bm{c}(\beta_i,\phi^{j},\gamma_k)}))$}

\BinaryInfC{$B_3\imply\ldots(B_n\imply p_{\bm{c}(\beta_i,\phi^{j},\gamma_k)})$}
\noLine
\UnaryInfC{$\vdots$}\noLine\UnaryInfC{$\vdots$}\noLine\UnaryInfC{$\vdots$}\noLine\UnaryInfC{$\vdots$}
\noLine
\UnaryInfC{$B_n\imply p_{\bm{c}(\beta_i,\phi^{j},\gamma_k)}$}
\insertBetweenHyps{\hskip -170pt}
\BinaryInfC{$p_{\bm{c}(\beta_i,\phi^{j},\gamma_k)}$}
\end{prooftree}

 The  derivation above shows the existence of a derivation of the translated conclusion from the translated premises.
\item[$\bullet$ Last rule is a $\bm{c_m}$-elim rule.] Then $\Pi^{\star}$ is as following:
\def\proofSkipAmount{\vskip .7cm}
\begin{prooftree}
\AxiomC{$\Pi_{pm}^{\star}$}
\noLine
\UnaryInfC{$\bm{c}(\beta_i,\phi^{j} )$}
\AxiomC{$\Pi_{i,j}^{\star}$}
\noLine
\UnaryInfC{$(\phi^{j}_i)_{i,j}$}
\AxiomC{$[\beta_1]$}
\noLine
\UnaryInfC{$\Pi_{1}^{\star}$}
\noLine
\UnaryInfC{$\chi$}
\AxiomC{$\ldots$}
\AxiomC{$[\beta_n]$}
\noLine
\UnaryInfC{$\Pi_n^{\star}$}
\noLine
\UnaryInfC{$\chi$}
\QuinaryInfC{$\chi$}
\end{prooftree}
By Proposition~\ref{sub-formula} $\Pi_{pm}^{\star}$, $\Pi_{i,j}^{\star}$ and $\Pi_i^{\star}$, $i=1,n$, $j=1,k$ satisfies the sub-formula principle. By inductive hypothesis, for each $i$ (and $j$), there are derivations $\Pi^{\prime}_{i}$, $\Pi^{\prime}_{i,j}$ and $\Pi_{pm}^{\prime}$  satisfying the statement of the lemma. Using, for each $i$ and $j$ the derivations $\Pi^{\prime}_{i,j}$ we obtain the following derivation from the 
$\epsilon$-axiom $\epsilon(r,\mathcal{M}(\chi),\mathcal{M}(\beta_i),\varphi_{i,j})$.
{\scriptsize
\begin{prooftree}
%\hskip -40pt
\AxiomC{$\Pi_{n,j_n}^{\prime}$}
\noLine
\UnaryInfC{$\mathcal{M}(\phi^{n}_{j_n})$}
\AxiomC{$\Pi_{1,2}^{\prime}$}
\noLine
\UnaryInfC{$\mathcal{M}(\phi^{1}_2)$}
\AxiomC{$\Pi_{1,1}^{\prime}$}
\noLine
\UnaryInfC{$\mathcal{M}(\phi^{1}_1)$}
\AxiomC{$\mathcal{M}(\phi^{1}_{1})\imply(\ldots(\mathcal{M}(\phi^{n}_{j_n})\imply(..(\mathcal{M}(\beta_1)\imply\mathcal{M}(\chi))\imply\ldots(\mathcal{M}(\beta_n)\imply\mathcal{M}(\chi))\imply(p_{\bm{c}(\beta_i,\phi^{j})}\imply\mathcal{M}(\chi)))))$}
\BinaryInfC{$\mathcal{M}(\phi^{1}_{2})\imply(\ldots(\mathcal{M}(\phi^{n}_{j_n})\imply(..(\mathcal{M}(\beta_1)\imply\mathcal{M}(\chi))\imply\ldots(\mathcal{M}(\beta_n)\imply\mathcal{M}(\chi))\imply(p_{\bm{c}(\beta_i,\phi^{j})}\imply\mathcal{M}(\chi)))))$}
%\doubleLine
\BinaryInfC{$\mathcal{M}(\phi^{1}_{3})\imply(\ldots(\mathcal{M}(\phi^{n}_{j_n})\imply(..(\mathcal{M}(\beta_1)\imply\mathcal{M}(\chi))\imply\ldots(\mathcal{M}(\beta_n)\imply\mathcal{M}(\chi))\imply(p_{\bm{c}(\beta_i,\phi^{j})}\imply\mathcal{M}(\chi)))$}
\noLine
\UnaryInfC{$\vdots$}\noLine\UnaryInfC{$\vdots$}\noLine\UnaryInfC{$\vdots$}\noLine\UnaryInfC{$\vdots$}
\noLine
\UnaryInfC{$\mathcal{M}(\phi^{n}_{j_n})\imply(..\mathcal{M}(\beta_1)\imply\mathcal{M}(\chi))\imply\ldots(\mathcal{M}(\beta_n)\imply\mathcal{M}(\chi))\imply(p_{\bm{c}(\beta_i,\phi^{j})}\imply\mathcal{M}(\chi)))$}
\insertBetweenHyps{\hskip -70pt}
\BinaryInfC{$(\mathcal{M}(\beta_1)\imply\mathcal{M}(\chi))\imply\ldots(\mathcal{M}(\beta_n)\imply\mathcal{M}(\chi))\imply(p_{\bm{c}(\beta_i,\phi^{j})}\imply\mathcal{M}(\chi)))$}
\end{prooftree}
}

We denote the derivation above by $\Sigma$. Using $\Pi^{\prime}_i$, for each $i$, we have the following derivation.
\def\proofSkipAmount{\vskip 1cm}
\begin{prooftree}
\AxiomC{$[\mathcal{M}(\beta_i)]$}
\noLine
\UnaryInfC{$\Pi_{i}^{\prime}$}
\noLine
\UnaryInfC{$\mathcal{M}(\chi)$}
\UnaryInfC{$\mathcal{M}(\beta_i)\imply \mathcal{M}(\chi)$}
\end{prooftree}

Then the  derivation below, built using the above derivations,  shows how to conclude the translation of the conclusion of $\Pi^{\star}$ from the translation of its premises and the implicational introduction and elimination schemata. $\epsilon(r,\mathcal{M}(\chi),\mathcal{M}(\beta_i),\varphi_{i,j})$ is the $\epsilon$-axiom that implements the implicational elimination on the implicational fragment of minimal logic. We remind the reader that $\mathcal{M}(\bm{c}(\beta_i,\phi^{j} ))$ is $p_{\bm{c}(\beta_i,\phi^{j})}$. We can check and find out that the derivation satisfies the lemma. 

{\scriptsize
\def\proofSkipAmount{\vskip 1cm}
\begin{prooftree}
\AxiomC{$\Pi_{pm}^{\prime}$}
\noLine
\UnaryInfC{$p_{\bm{c}(\beta_i,\phi^{j})}$}
\AxiomC{$[\mathcal{M}(\beta_n)]$}
\noLine
\UnaryInfC{$\Pi_{n}^{\prime}$}
\noLine
\UnaryInfC{$\mathcal{M}(\beta_n)\imply \mathcal{M}(\chi)$}
\AxiomC{$[\mathcal{M}(\beta_1)]$}
\noLine
\UnaryInfC{$\Pi_{1}^{\prime}$}
\noLine
\UnaryInfC{$\mathcal{M}(\beta_1)\imply \mathcal{M}(\chi)$}
\AxiomC{$\Sigma$}
\noLine
\UnaryInfC{$(\mathcal{M}(\beta_1)\imply\mathcal{M}(\chi))\imply\ldots(\mathcal{M}(\beta_n)\imply\mathcal{M}(\chi))\imply(p_{\bm{c}(\beta_i,\phi^{j})}\imply\mathcal{M}(\chi))$}
\BinaryInfC{$(\mathcal{M}(\beta_2)\imply\mathcal{M}(\chi))\imply\ldots(\mathcal{M}(\beta_n)\imply\mathcal{M}(\chi))\imply(p_{\bm{c}(\beta_i,\phi^{j})}\imply\mathcal{M}(\chi))$}
\noLine
\UnaryInfC{$\vdots$}
\noLine
\UnaryInfC{$\vdots$}
\noLine
\UnaryInfC{$(\mathcal{M}(\beta_n)\imply\mathcal{M}(\chi))\imply(p_{\bm{c}(\beta_i,\phi^{j})}\imply\mathcal{M}(\chi))$}
\insertBetweenHyps{\hskip -70pt}
\BinaryInfC{$p_{\bm{c}(\beta_i,\phi^{j})}\imply\mathcal{M}(\chi)$}
\BinaryInfC{$\mathcal{M}(\chi)$}
\end{prooftree}
}

\end{description}

\begin{center}
Q.E.D.
\end{center}

For logics that have rules with a proviso, the respective implicational axioms $\iota$ and $\epsilon$ must reflect the corresponding conditions. Whenever we have these axioms we have Lemma~\ref{lemma1} holding for the respective logics. For example, this can be done with the Natural Deduction rules provided in \cite{Pra:ND} for S4 and S5 and already discussed before in this article.   

%We now can provide a translation that preserves theorems in $\mathcal{L}$ as direct consequence of lemma~\ref{lemma1}

\begin{proposition}
Let $\mathcal{L}$ be a propositional logic satisfying the sub-formula principle. Consider the following translation $\star$ from formulas of \Log{L} into formulas of \mil:
Let  $\mathcal{A}^{\alpha}(\alpha)$ be $\{\varphi_1,\ldots,\varphi_k\}$  
$\alpha^{\star}$ is defined as $\varphi_1\imply (\varphi_2\imply \ldots (\varphi_k\imply \mathcal{M}(\alpha)))$. 
Thus, $\vdash_{\mathcal{L}}\alpha$ if and only if $\vdash_{\mil}\alpha^{\star}$.
\end{proposition}

{\em Proof of proposition}. The proposition follows immediately from Lemma~\ref{lemma1}. 

\begin{center}
Q.E.D.
\end{center}

\begin{proposition}
If $\alpha$ is any propositional formula, then the number of sub-formulas of $\alpha$ is polynomially bounded on the length of $\alpha$.
\end{proposition}

{\em Proof}. Each sub-formula of $\alpha$ is determined by the main connective of it. In $\alpha$ there is at most one connective or logical constant by symbol position in $\alpha$. Thus, the number of sub-formulas of $\alpha$ is bounded by the length of $\alpha$.

We can see that the size of $\alpha^{\star}$ is $O(m^{3})$, if $m$ is the size of $\alpha$. The formula $\alpha^{\star}$ depends on three choices  of sub-formulas of $\alpha$. This polynomial bound on the size of $\alpha^{\star}$ entails the following main conclusions of our article.  

\begin{theorem}\label{principal}
If $\mathcal{L}$ satisfies the sub-formula principle then the problem of knowing whether $\alpha$, a formula of $\mathcal{L}$, is provable or not, is in PSPACE. If $\mathcal{L}$ includes \mil~ then this problem, also known as {\bf Validity}, is PSPACE-complete. 
\end{theorem}

{\em Proof of theorem}. From the PSPACE-completeness of provability in \mil and the polynomial reduction of L $\mathcal{L}$ to \mil, we have that {\bf Validity} in $\mathcal{L}$, namely, knowing whether $\alpha$, a formula of $\mathcal{L}$, is provable or not, is in PSPACE. If $\mathcal{L}$ includes \mil, and since \mil is PSPACE-complete,  then {\bf Validity} is PSPACE-complete. 

\begin{center}
Q.E.D.
\end{center}

% in the case that {\bf Validity} is PSPACE-complete and CoPSPACE-complete are the same class, and so% {\bf Invalidity}. or better {\bf Unprovability} is $PSPACE$-complete too. 

\begin{corollary}\label{corolario}
If $\mathcal{L}$ satisfies the sub-formula principle and includes \mil then the problem of knowing whether $\alpha$, a formula of $\mathcal{L}$, is invalid or not, is PSPACE-complete.
\end{corollary}

{\em Proof of corollary}. Since $PSPACE$ is a deterministic class, we have that CoPSPACE=PSPACE. The problem stated in the corollary is in CoPSPACE, by definition. 

\begin{center}
Q.E.D.
\end{center}

\section{A brief discussion  on extending the results to finitely many-valued propositional logics}

An  stronger version of the statement of Theorem~\ref{principal} can be used to prove that some finitely many-valued logics have {\bf Validity} (cf. Theorem~\ref{principal}) in PSPACE. In \cite{IGPLCica} it is provided a general schema to define normalizable Natural Deduction systems for some finitely many-valued logics. The kind of Natural Deduction rules used in \cite{IGPLCica} are similar, in fact almost the same, to those shown here in this article. However, when formalizing logics with more than 2 truth-values, auxiliary formulas are considered in order to perform the {\em binary print}, explained in sequel, of a truth-value. We briefly discuss here the main idea on an extension of Theorem~\ref{principal} and how it could be used to prove that Validity/Provability in a finitely many-valued logic is in PSPACE, by adjusting the sub-formula principle defined in Definition~\ref{SFP}. More details on the Natural Deduction systems for finitely many-valued logics as used here can be found in \cite{IGPLCica}. 

Instead of defining a rule for each of the $n>2$ truth-values, it is used a method to reduce many-valued semantics to bivalent one. We use the three-valued logic $\text{\L}_{3}$, due to \L ukasiewicz to illustrate our discussion. See \cite{L3} for the original article (1920) on $\text{\L}_3$ and \cite{Malinowski} for a very good and concise presentation of many-valued logics. 

The truth-values in $\text{\L}_3$ are $\{0,i,1\}$, where 0 and $i$ are undesignated values and $1$ is the designated value. With the sake of providing some context on the meaning of these three values, we can say that the truth-value $i$ means indetermination or in numerical terms that it corresponds\footnote{This depends whether we are in an epistemic position, the former,  or an ontic position, the later case.} to 1/2. Thus,  a formula having $i$ as value should not be taken as a true formula. The truth-values $i$ and $0$ in this case are considered as undesignated. 
For some many-valued logics, a formula is valid, iff, it yields designated value for each possible truth-value assignment. This is just the case with $\text{\L}_3$. $\text{\L}_3$ has the truth-table shown in Table~\ref{L3} for the implication $\imply$ and in Table~\ref{L3neg} for the negation $\neg$. We explain what is a {\em bitprint} of a truth-value in the sequel. Consider a function $t(P)$ that yields 1 if $P$ is designated and $0$ if it is undesignated. Using the formula $\phi(P)$, defined as $\neg P\imply P$, it is possible to identify $i$ with the pair $\langle 0,1\rangle$, 1 with the pair $\langle 1,1\rangle$ and 0 with $\langle 0,0\rangle$, that is, each truth-value $x$ is identified by the pair $\langle t(x),\phi(x)\rangle$.  
 In \cite{JoaoCaleiro}, the sequence $\langle P,\phi(P)\rangle$ that denotes each one of the values in $\{0,i,1\}$ is called the bitprint of the respective truth-value. It is  interesting to note that the pair $\langle 1,0\rangle$ has no truth-value associated to it. Besides that, it is worth observing that for a many-valued logic with $k$ truth-values, we have to find $maxint(log_{2}(k))-1$ separating formulas for performing the role of $\phi(P)$. Using bitprints, the truth-table of the $\limply$ is as shown in Table~\ref{L3bitprint}.
In this way the line of the truth-table saying that $P\limply Q$ yields 1, when 
$P$ is $i$ and $Q$ is 1, is coded as: $P$ is 0, $\phi(P)$ is 1, $Q$ is 1 and $\phi(Q)$ is 1. This line, the fourth line of the truth-table, is related to the Natural Deduction rule shown in Figure~\ref{L3Imp}. The rule in Figure~\ref{ElimL3}, very similar to an elimination rule in Natural Deduction, corresponds to the third line in Table~\ref{L3bitprint}. Because some pairs $\langle t(P),\phi(P)\rangle$ do not correspond to any truth-value, there is need of a Natural Deduction rule to take care of this. The rule that is related to the pair $\langle 1,0\rangle$ in $\text{\L}_3$ is shown in Figure~\ref{L3OtherValues}. Besides that, rules as shown in Figure~\ref{PhiRules}, are need in order to have a complete system, for, in the general case, formulas of the form $\phi(X)$, may have $X$ as composed formulas. Figure~\ref{PhiRules} shows the case when $X$ is $P\imply Q$. This is the case related to the first line of the truth-table in Figure~\ref{L3bitprint}. 

\begin{figure}[!htb]
\centering
\begin{minipage}[b]{0.45\linewidth}
\[
 \begin{array}{|c|c|c|}
   \hline
P & Q & P\limply Q\\
\hline
1&1&1\\
\hline
1&i&i\\
\hline
1&0&0\\
\hline
i&1&1\\
\hline
i&i&1\\
\hline
i&0&i\\
\hline
0&1&1\\
\hline
0&i&1\\
\hline
0&0&1\\
\hline
  \end{array}
\]
\caption{Truth-table for $\limply$ in $\text{\L}_3$}
\label{L3}
\end{minipage}
%\end{figure}
\quad
%\begin{figure}[!htb]
%\centering
\begin{minipage}[b]{0.45\linewidth}
\[
 \begin{array}{|c|c|}
   \hline
P & \neg P\\
\hline
1&0\\
\hline
i&i\\
\hline
0&1\\
\hline
  \end{array}
\]
\caption{Truth-table for $\neg$ in $\text{\L}_3$}
\label{L3neg}
\end{minipage}
\end{figure}

\begin{figure}[!htb]
\[
 \begin{array}{|c|c||c||c|c|c|c||c|c||}
   \hline
P & Q & P\limply Q & t(P) & \phi(P) & t(Q) & \phi(Q) & t(P\limply Q) & \phi(P\limply Q) \\
\hline
1&1&1&1&1&1&1&1&1\\
\hline
1&i&i&1&1&0&1&0&1\\
\hline
1&0&0&1&1&0&0&0&0\\
\hline
i&1&1&0&1&1&1&1&1\\
\hline
i&i&1&0&1&0&1&1&1\\
\hline
i&0&i&0&1&0&0&0&1\\
\hline
0&1&1&0&0&1&1&1&1\\
\hline
0&i&1&0&0&0&1&1&1\\
\hline
0&0&1&0&0&0&0&1&1\\
\hline
  \end{array}
\]
\caption{Truth-table for $\limply$ in $\text{\L}_3$ using bitprints in the 6 last columns}
\label{L3bitprint}
\end{figure}

\begin{figure}[!htb]
\begin{prooftree}
\AxiomC{$[P\limply Q]$}
\noLine
\UnaryInfC{$\vdots$}
\noLine
\UnaryInfC{$C$}
\AxiomC{$[P]$}
\noLine
\UnaryInfC{$\vdots$}
\noLine
\UnaryInfC{$C$}
\AxiomC{$\phi(P)$}
\AxiomC{$Q$}
\AxiomC{$\phi(Q)$}
\QuinaryInfC{$C$}
\end{prooftree}
\caption{$\limply$-introduction rule for the logic $\text{\L}_3$ related to the fourth line in Figure~\ref{L3bitprint}}
\label{L3Imp}
\end{figure}

\begin{figure}[h]
\begin{prooftree}
\AxiomC{$[P]$}
\noLine
\UnaryInfC{$\vdots$}
\noLine
\UnaryInfC{$C$}
\AxiomC{$[\phi(P)]$}
\noLine
\UnaryInfC{$\vdots$}
\noLine
\UnaryInfC{$C$}
\AxiomC{$Q$}
\AxiomC{$\phi(Q)$}
\AxiomC{$P\limply Q$}
\QuinaryInfC{$C$}
\end{prooftree}
\caption{$\limply$-elimination-like rule for the logic $\text{\L}_3$ related to the third line in Figure~\ref{L3bitprint}}
\label{ElimL3}
\end{figure}

\begin{figure}[h]
\begin{prooftree}
\AxiomC{$P$}
\AxiomC{$[\phi(P)]$}
\noLine
\UnaryInfC{$\vdots$}
\noLine
\UnaryInfC{$C$}
\BinaryInfC{$C$}
\end{prooftree}
\caption{$\limply$-rule for the logic $\text{\L}_3$ related to the value $\langle 1,0\rangle$ that is not a bitprint}
\label{L3OtherValues}
\end{figure}

\begin{figure}[!htb]
\begin{prooftree}
\AxiomC{$[P]$}
\noLine
\UnaryInfC{$\vdots$}
\noLine
\UnaryInfC{$C$}
\AxiomC{$[\phi(P)]$}
\noLine
\UnaryInfC{$\vdots$}
\noLine
\UnaryInfC{$C$}
\AxiomC{$[Q]$}
\noLine
\UnaryInfC{$\vdots$}
\noLine
\UnaryInfC{$C$}
\AxiomC{$[\phi(Q)]$}
\noLine
\UnaryInfC{$\vdots$}
\noLine
\UnaryInfC{$C$}
\AxiomC{$[\phi(P\limply Q)]$}
\noLine
\UnaryInfC{$\vdots$}
\noLine
\UnaryInfC{$C$}
\QuinaryInfC{$C$}
\end{prooftree}
\caption{$\phi(\limply)$-intro rule in $\text{\L}_3$ regard to $\phi(P\limply Q)$ and first line in Figure~\ref{L3bitprint}}
\label{PhiRules}
\end{figure}

Consider a general propositional connective $c$, that forms formulas of the form $c(A_1,\ldots,A_n)$. The rules that discharge $c(A_1,\ldots,A_n)$ are considered $c$-introduction rules. The rules that have $c(A_1,\ldots,A_n)$ as premise (also called major premise) are the $c$-elimination rules. The same is said about $\phi(c(A_1,\ldots,A_n))$, where $\phi$ is a separating formula. In \cite{IGPLCica} it is shown how to eliminate sequences that starts with $c$-intro rules and ends with $c$-elimination rules. Derivations having this kind of sequences are said to be non-normal. Normal derivations do not have these sequences, and, do not have $c$-rules, like the one shown in Figure~\ref{L3OtherValues}, proving premise of $c$-elimination rules either. It is the case that any non-normal derivation of a formula $\alpha$ from a set of formulas $\Gamma$ can be effectively transformed in a normal derivation of $\alpha$ from $\Gamma^{\prime}\subseteq\Gamma$. This is the statement of the normalization theorem for finitely many-valued logics proved in \cite{IGPLCica}.

 A branch in a derivation in a Natural Deduction system for a finitely many-valued logic, as those discussed in this section,  is any sequence of formulas starting in an undischarged hypothesis  and ending in the minor premise of an elimination rule. This is essentially what is stated in Definition~\ref{branch}. One of the properties of normal derivations is that for any branch starting with a formula $A$ and ending with $B$ in this derivation,  the formulas that occur in it are either sub-formulas of $A$, or $B$, or both. This is one of the main features of normal proofs, in usual systems as Classical, Intuitionistic and Minimal logic. As was already mentioned in the introduction, normalization implies in the sub-formula property in many logics. A quick inspection on the form of the general Natural Deduction introduction and elimination rules as defined in Section~\ref{translation} shows that for these rules the premises are either sub-formula of the conclusion or vice-versa. The discharged formulas of both, introduction and elimination are also sub-formula either of the major premise or of the conclusion. In fact, the sub-formula principle holds because of this feature just discussed. However, the format of the Natural Deduction rules for the finitely many-valued logics does not show this perfect relationship. In their case, the use of the separating formulas $\phi(X)$ disturbs this relationship based only on sub-formula occurrence in rules. In order to apply our technique to finitely many-valued logic, we will consider the sub-formula relationship through application of any formula in the set of separating formulas $\{\phi_1,\ldots,\phi_n\}$, so that the relationship between conclusions , premises and discharged formulas, in this case of Natural Deduction rule, includes the separating formulas. For example, $P$, the discharged formula, and $Q$, a minor premise, are sub-formulas of the major premise $P\limply Q$, of the rule in Figure~\ref{ElimL3}. However, $\phi(P)$, a discharged formula, and $\phi(Q)$, other minor premise, are not sub-formulas. They are, in fact, images of sub-formulas of the major premise. So, we can say that the formulas occurrences in a Natural Deduction rule, as in Figure~\ref{ElimL3}, are $\phi$-related. This is defined below. 

\begin{definition}
Consider a Natural Deduction rule in a system for finitely many-valued logic using a set $\Phi=\{\phi_1,\dots,\phi_n\}$ of separating formulas. The relationship between premises, conclusions and discharged formulas under sub-formula of $\phi_i$, $i=1,n$, image of a sub-formula is denoted by $\Phi$-sub-formula relationship.
\end{definition}

It is a trivial fact that if $\Phi\subseteq\Phi^{\prime}$ and $\alpha$ is $\Phi$-sub-formula of $\beta$, or vice-versa, then they are $\Phi^{\prime}$-sub-formula too. From the form of a branch beginning with $\beta_1$ and ending in $\beta_k$ in a normal derivation we can see that any formula in this branch is either a sub-formula of $\beta_1$ or of $\beta_n$, and for any $i=1,n-1$, $\beta_i$ is $\Phi$-sub-formula of $\beta_{i+1}$, or vice-versa. Using this, we can use the same  polynomial simulation used for proving Theorem~\ref{principal} for any finitely many-valued logic in $\mil$. It is worth noting that the conclusion and hypothesis of any normal derivation impose that the set of formulas that occur in it are either sub-formulas of the conclusion or of the hypothesis. Does not matter the fact that some of then are $\Phi$-images of some other sub-formula.

As a matter of illustration of the discussion in the last paragraph, we show the implicational schemata for the elimination, introduction and $\phi$-rules, including the $\phi$-rules related to no bitprint representation (as the one shown in Figure~\ref{L3OtherValues}). For example, in the case of $\text{\L}_3$, the implicational schema for the rule shown in Figure~\ref{L3Imp} is below, where $C$ is a sub-formula of either the conclusion $\alpha$ of the derivation, or of some hypothesis of the derivation. There must be one such formula for each $C$, this already includes the images under the separating formula $\phi$. We remember that $p_{(P\limply Q)}$ is the new (fresh) propositional variable associated to $P\limply Q$. The same holds for the propositional variables $p_{\phi(P)}$ and $p_{\phi(Q)}$
\[
(p_{(P\limply Q)}\to C)\to ((P\to C)\to (p_{\phi(P)}\to (Q\to (p_{\phi(Q)}\to C))))
\]  
The formula below is the implicational schema for the corresponding $\phi$-introduction rule in Figure~\ref{PhiRules}.
\[
(P\to C)\to ((p_{\phi(P)}\to C)\to ((Q\to C)\to ((p_{\phi(Q)}\to C)\to ((p_{(P\limply Q)}\to C)\to C))))
\]
The implicational schema for the $\limply$-elimination rule (Figure~\ref{ElimL3}) and the rule shown in Figure~\ref{PhiRules} are show in the sequel. 
\[
(P\to C)\to ((p_{\phi(P)}\to C)\to (Q\to (p_{\phi(Q)}\to (p_{(P\limply Q)}\to C))))
\]  
\\
\[
P\to ((p_{\phi(P)}\to C)\to C)
\]

{\bf Important Observation}: The reader must have noted that the forms of the implicational schemata are not totally in formal agreement with the $\iota$-axioms presented in the previous sections and used to prove that {\bf Validity} of the respective logic is in PSPACE. This is only a formal disagreement, indeed. This can be seen, if we consider the usual conjunction $\land$. The $\land$ very well-known introduction rules is the left rule below, but it can be also defined as drawn by the right rule. 
\begin{prooftree}
\AxiomC{$A$}
\AxiomC{$B$}
\BinaryInfC{$A\land B$}
\AxiomC{$A$}
\AxiomC{$B$}
\AxiomC{$[A\wedge B]$}
\noLine
\UnaryInfC{$\vdots$}
\noLine
\UnaryInfC{$C$}
\TrinaryInfC{$C$}
\noLine 
\BinaryInfC{.}
\end{prooftree}

Finally, from the proof-theoretical results obtained in \cite{IGPLCica} we can state the following theorem. A detailed proof of this theorem and a more systematic treatment of the subject discussed in this section is omitted due to the lack of space. 

\begin{theorem}
Let \Log{L} be a propositional $k$-valued logic with finitely-many connectives with semantics determined by functional truth-tables. The {\bf Validity} decision problem for \Log{L} is in PSPACE.  
\end{theorem}

{\bf Proof}. The fact that the connectives are defined by functional truth-tables, i.e. deterministic and defined for every $k$-value truth-tables, entails that the Natural Deduction system is sound and complete regarding the semantics provided by the truth-tables. Consider the set $\{\phi_1,\ldots,\phi_{log_2(k)}\}$ of separating formulas that provide bitprints for each of the $k$ truth-values. There is at most one rule for each line in each truth-value. Since the number of lines in each truth-table is fixed (depends only on $k$), there is at most one implicational schema for each line. Let $F(k)$ be this number. $F(k)$ depends only on $k$. In fact $F(k)$ is induced by the arity of each connective of the language of \Log{L}. Taking into account a formula $\alpha$ in the language of \Log{L}.
Each 
implicational schema depends also on the number of sub-formulas of $\alpha$, including those sub-formulas that are image under each $\phi_i$. Let us say that $n$ is the number of sub-formulas of the formula $\alpha$, that is linear on the size of $\alpha$. Thus, the total number of implicational schemata is at most $n^{a\star log_{2}(k)}$, where $a=F(k)$. This upper-bound is polynomial in $n$. Thus, the translation used in Theorem~\ref{principal} can the used here to prove that {\bf Validity} is in PSPACE. It is worth noting that $F(k)$ depends only on the logic. 
\begin{center}
{\bf Q.E.D.}
\end{center}

If the Validity of a  logic \Log{L} is in PSPACE  and it includes \mil~ then its (Validity decision problem) is PSPACE-complete. A logic $\Log{L}$ includes \mil, iff, the $\imply$-intro and $\imply$-elim are derived rules in \Log{L}.

\section{Conclusion}
%\section{Discussion}
\label{applications}

We have shown that structural propositional logics satisfying the sub-formula principle have their provability problem in PSPACE and that if they include \mil then they are PSPACE-complete problems. We used proof-theory to show this. An important consequence of our result is to show that some logics hardly satisfy the sub-formula principle in the terms here presented. Any propositional logic that we believe that it is beyond PSPACE cannot satisfy the sub-formula principle.  

An immediate application of Theorem~\ref{principal} shows that hardly some well-known logics, as Propositional Dynamic Logic ($PDL$, see \cite{PDL}) and the common-knowledge epistemic logic with at least 2 agents \cite{CommonEpistemicLogic}, have Natural Deduction systems satisfying the sub-formula principle. In \cite{ND-PDL}, for example, it is shown a Natural Deduction system for PDL, with normalization theorem. However, one of its rules, the iteration rule, does not satisfy the sub-formula principle in the terms stated in this article, since it is an infinitary ND-system. Anyway it does not satisfy sub-formula relationship between premises and conclusion either. Any logic that is EXPTIME-complete, as the ones just discussed,  cannot satisfy the sub-formula principle too. 

Any PSPACE-complete problem is NP-Hard. Since Classical propositional Logic can be specified using the general elimination and introduction rules presented here, then it is PSPACE-HARD and so its also NP-Hard. We know, by the truth-table method that Classical propositional logic is in NP, so it is NP-complete. 

We discussed the extension of this proof-theoretical way of providing upper-bounds for provability to finitely many-valued logics, concluding that their respective provability (validity) problems are in PSPACE. This is in our opinion the far we can go. The fact that the set of valid formulas of a logic is in PSPACE may be of no help in some particular analysis. There are many-valued logics, as for example Kleene's logic, that have no tautologies and this is of course trivially in PSPACE. On the other hand, knowing that many-valued usual logics are in PSPACE seems to be an interesting fact in general. There are finitely many-valued logics that have non-trivial set of tautologies.   As far as we know there should be no novelty in proving that Validity in finitely many-valued logics is in PSPACE. A polynomially space-bounded algorithm for testing Validity can be designed by observing that evaluating the truth-value of a formula is polynomial on the size of the formula and that the evaluation of other rows can reuse the space. Here we presented an alternative proof of the fact that Validity is in PSPACE. The interesting part is to conclude that if the logic includes \mil~, then its Validity is PSPACE-complete.

\subsection{Further investigation }

We have to consider the known PSPACE-complete propositional logics. The best well-known are the Modal Logics $K$, $S4$, and $KD$.We know that (unlabeled) Natural Deduction systems for these logics are problematic. They fail to satisfy normalization and have rules with complex provisos. A consequence of this is that the natural reductions expected to work according the inversion principle produce derivations out of the system. Thus, the application of our technique to these logics depends on a concrete Natural Deduction system with rules that are instances of the general rules we deal with. Anyway, we already know, by other methods that these logics are PSPACE-complete, so this application would not be worth of presenting. However, there are labeled Natural Deduction systems with normalization theorem and a related  sub-formula lemma. 

For logics that we do not know the complexity class of the provability problem, the extension of the technique presented in this article to the a kind of general labeled introduction and elimination system of rules would be worth of developing. 
Our general form for introduction and elimination rules does not consider the many approaches of labeled Natural Deduction. It is our intention to discuss how Theorem~\ref{principal} can be also obtained for the usual labeled systems used to specify modal logics, namely, \cite{MartiniMasini}, \cite{Simpson} and finally \cite{BasinMathewsVigano}. This would be quite useful in providing a proof-theoretical complexity analysis for the intuitionistic versions of Modal Logics.

Extending this technique to first-order logics would be interesting too. There are decidable fragments of pure predicate (without functions and the equality ``='') that are beyond PSPACE, and surely the statement of the sub-formula principle for these cases are worth of studying. 
Finally, investigations concerning the feasibility of the interpolant (see \cite{Pitasi}) and the existence of ND provers with sub-formula property are worth of studying too.

%\section{Acknowledgements}
{\em Acknowledgments:}
We thank many colleagues for suggesting improvements in the work reported in this article. Particularly, Luiz Carlos Pereira, Lew Gordeev, Mario Benevides, Wagner Sanz and Peter Schroeder-Heister, Cecília Englander, Bruno Lopes and Jefferson de Barros Santos. We would like to mention Carlos Caleiro and Jo\~{a}o Marcos for pointing out that finitely many-valued logics could be a better example of positive application of our proof-theoretical technique than Modal Logics. We are very glad and thankful to have followed their suggestions.

\bibliographystyle{eptcs}
\bibliography{implicational}

\end{document}